\documentclass{emulateapj}
\usepackage{amssymb, amsmath, apjfonts, natbib, color}
\def\degree{{\circ}}
\newdimen\digitwidth
\setbox1=\hbox{0}
\digitwidth=\wd1
\catcode`"=\active
\def"{\kern\digitwidth}

\begin{document}
\title{UNCERTAINTIES IN THE DEPROJECTION OF THE OBSERVED BAR PROPERTIES}

\author{Yanfei Zou\altaffilmark{1}, Juntai Shen\altaffilmark{1, 2} and Zhao-Yu Li\altaffilmark{1}}

\altaffiltext{1}{Key Laboratory for Research in Galaxies and Cosmology, Shanghai Astronomical Observatory, Chinese Academy of Sciences, 80 Nandan Road, Shanghai 200030, China}
\altaffiltext{2}{Correspondence should be addressed to Juntai Shen: jshen@shao.ac.cn}

\begin{abstract}
In observations, it is important to deproject the two fundamental quantities characterizing a bar, i.e., its length ($a$) and ellipticity ($e$), to face-on values before any careful analyses. 
However, systematic estimation on the uncertainties of the commonly used deprojection methods is still lacking. 
Simulated galaxies are well suited in this study. 
We project two simulated barred galaxies onto a 2D plane with different bar orientations and disk inclination angles ($i$). 
Bar properties are measured and deprojected with the popular deprojection methods in the literature. 
Generally speaking, deprojection uncertainties increase with increasing $i$. 
All the deprojection methods behave badly when $i$ is larger than $60^\degree$, due to vertical thickness of the bar. 
Thus, future statistical studies of barred galaxies should exclude galaxies more inclined than $60^\degree$. 
At moderate inclination angles ($i\leq60^\degree$), 2D deprojection methods (analytical and image stretching) and Fourier-based methods (Fourier decomposition and bar-interbar contrast) perform reasonably well with uncertainties $\sim10\%$ in both the bar length and ellipticity. 
Whereas the uncertainties of the 1D analytical deprojection can be as high as 100\% in certain extreme case.

We find that different bar measurement methods show systematic differences in the deprojection uncertainties. 
We further discuss the deprojection uncertainty factors with the emphasis on the most important one, i.e., the 3D structure of the bar itself. 
We construct two triaxial toy bar models that can qualitatively reproduce the results of the 1D and 2D analytical deprojections; they confirm that the vertical thickness of the bar is the main source of uncertainties.

\end{abstract}


\section{INTRODUCTION}

Bars are commonly seen in the disk galaxies and play an important role in the secular evolution (\citealt{1993A&A...268...65F}; \citealt{2005ApJ...632..217S}; \citealt{2011MNRAS.411.2026M}; \citealt{2011MNRAS.415.3308G}; see \citealt{2004ARA&A..42..603K} for review). 
Optical studies found that the bar fraction is typically around 50\% (\citealt{2007ApJ...659.1176M}; \citealt{2008ApJ...675.1194B}; \citealt{2009A&A...495..491A}). 
In the near-infrared (NIR) images, where the galaxy morphology is not strongly influenced by dust extinction and star forming regions, the bar fraction can be as high as $\sim$65\% (\citealt{2000AJ....119..536E}; \citealt{2007ApJ...657..790M}). 
Bars are also found in the high redshift disk galaxies with possible cosmological evolution on the bar fraction (\citealt{2004ApJ...615L.105J}; \citealt{2008ApJ...675.1194B}; \citealt{2008ApJ...675.1141S, 2012ApJ...758..136S}).

By effectively redistributing angular momentum and energy of disk material, bars drive the morphological evolution of disk galaxies secularly (\citealt{1985MNRAS.213..451W}; \citealt{1998ApJ...493L...5D}; \citealt{2003MNRAS.341.1179A}). 
The large-scale mixing of the interstellar gas driven by the bar can flatten the chemical distribution of the disk (\citealt{1994ApJ...424..599M}; \citealt{2001AJ....122.1298G}). 
Bars might also help fuel the central supermassive black hole (SMBH) to trigger the active galactic nucleus (AGN) activity (\citealt{1989Natur.338...45S}; \citealt{1997ApJ...487..591H}; \citealt{2011ApJ...743L..13C}). 
However, recent statistical studies on the connection between bars and AGN activities found controversial results (\citealt{2009ASPC..419..402H}; \citealt{2012ApJ...745..125L}; \citealt{hao..et..al..2014}). 
Moreover, in simulations, once a bar has formed, its inner part quickly buckles in the vertical direction, producing the observed boxy/peanut-shaped bulges (\citealt{1981A&A....96..164C}; \citealt{1991Natur.352..411R}; \citealt{1999ApJ...522..686B}).

Bars have different strengths in different galaxies. 
Bar strength is suggested to be connected with bulges, galaxy types, dark matter halos and gas transportation processes (\citealt{1995AJ....109.2428M}; \citealt{1985ApJ...288..438E}; \citealt{2003MNRAS.341.1179A}; \citealt{2012ApJ...758...14K}). 
There are several methods to measure the bar strength in the literature, such as the ellipticity of the bar (\citealt{1995AJ....109.2428M}), and the bar-interbar contrast (\citealt{1985ApJ...288..438E}; \citealt{1996AJ....111.2233E}). 
\cite{2001ApJ...550..243B} suggested that the maximum value ($Q_{\rm b}$) of the ratio between the tangential force and the mean axisymmetric radial force in a barred disk galaxy can also be used to characterize the bar strength. 
This parameter gives us a direct impression to the actual force due to a bar.
Observationally, stronger bars are usually longer with higher ellipticities (\citealt{1995AJ....109.2428M}; \citealt{2007ApJ...657..790M}; \citealt{2011MNRAS.415.3308G}).
In this paper, we focus on two fundamental observed quantities characterizing the bar strength, i.e.,  bar length and ellipticity, which can be easily measured from the galaxy image.
Besides the simple visual estimation of the bar properties (\citealt{1979ApJ...227..714K}; \citealt{1995AJ....109.2428M}), commonly used bar identification and measurement approaches include the maximum or minimum of the bar ellipticity, the radial variation of the isophotal position angle, the radial profile of the phase angle and the relative amplitude of the Fourier $m = 2$ mode, and the bar-interbar contrast (\citealt{2002ApJ...567...97L}; \citealt{2003MNRAS.338..465A}; \citealt{2003ApJ...592L..13S}; \citealt{2005MNRAS.364..283E}; \citealt{2007ApJ...659.1176M}; \citealt{2007ApJ...657..790M}; \citealt{2011ApJS..197...22L}).

Since real galaxies are all inclined to certain extent, the measured bar parameters should be first deprojected to their corresponding face-on values. 
The basic assumption of the deprojection is that the outer part of a bar is infinitely thin (\citealt{2007MNRAS.381..943G}). 
However, real bars are usually thickened via the buckling instability, showing a boxy/peanut shaped bulge in the inner region (\citealt{1981A&A....99..362S}; \citealt{1981A&A....96..164C}; \citealt{1991Natur.352..411R}; \citealt{1999AJ....118..126B}; \citealt{2006MNRAS.370..753B}).
As a good example, the Milky Way also harbors a buckled bar with the boxy bulge in the central region (\citealt{1991ApJ...379..631B}; \citealt{2002MNRAS.330..591B}; \citealt{2005MNRAS.358.1309B}; \citealt{2007MNRAS.378.1064R}; \citealt{2013MNRAS.434..595C}).
The vertically thick boxy bulge, as evident in COBE images (\citealt{1994ApJ...425L..81W}; \citealt{1995ApJ...445..716D}), may simply be the Galactic bar viewed edge-on with the major axis tilted $20^\degree$ away from the Sun-Galactic center (GC) line (e.g. \citealt{2010ApJ...720L..72S}).
It even presents a vertical X-shaped structure related to the buckling process (\citealt{2010ApJ...721L..28N}; \citealt{2010ApJ...724.1491M}; \citealt{2012ApJ...757L...7L}). Thus, this basic assumption needs to be treated with caution when deprojecting bars.

In the 1D approximation, a bar is treated as a straight line segment which can be easily deprojected (\citealt{1995AJ....109.2428M}). 
This approximation is, of course, too simple for real bars. 
Assuming that the bar can be described by a planar ellipse, \cite{2007MNRAS.381..943G} provided a more sophisticated method to analytically deproject the bar. 
They also tried to directly stretch the inclined galaxy image to the face-on one with the total flux conserved. 
After comparing different bar deprojection methods, they concluded that, when the inclination angle is less than $50^\degree$, all the methods agree very well with each other within 20\% difference. 
However, for the inclined galaxies in their sample, the face-on bar properties are actually unknown. 
Moreover, the estimated inclination angles introduce additional uncertainties. 
As a matter of fact, several popular deprojection methods are widely used in observations to characterize the bar structures (\citealt{1995AJ....109.2428M}; \citealt{2007ApJ...659.1176M}; \citealt{2007MNRAS.381..943G}; \citealt{2011ApJS..197...22L}; \citealt{2000A&A...361..841A}).
The majority of these studies ignore the uncertainties in the deprojection process since it is impossible to know the true face-on values of the bar from observations.
Consequently, it is hard to know to what extent these deprojection methods are accurate.

The uncertainties in various deprojection methods can be best assessed by analyzing a simulated disk galaxy.
The true face-on bar properties of a simulated galaxy can always be measured readily.
A simulated galaxy can also be ``observed'' from different viewing angles.
The bar properties measured in inclined systems are then deprojected to the face-on values using different deprojection methods.
Thus, the uncertainties of the deprojection methods can be tested by comparing the deprojected bar properties to the true face-on values.
If inclination angles are treated as known quantities, the uncertainties are mainly from deprojection methods.
By projecting the 3D simulations onto a 2D plane from different viewing angles, we create mock images and apply different bar measurement and deprojection methods. 
The creation of mock images and bar measurements are described in Section 2. 
Section 3 presents different deprojection methods and the corresponding results, which are further discussed in Section 4 and finally concluded in Section 5.

\section{IMAGE CREATION AND BAR MEASUREMENT}

\subsection{Mock Images}

For the purpose of this work we study two $N$-body disk simulations here (Model A and Model B) shown in Figure~1.
Model A is taken from the simulation in \cite{2010ApJ...720L..72S}, which well matches the BRAVA stellar kinematics in the Milky Way bulge region. 
This simple $N$-body model simulates a disk galaxy with $10^{6}$ particles evolving in a rigid dark matter halo potential. 
Initially the disk is dynamically cold (Toomre's $Q\sim 1.2$). 
A bar forms spontaneously and quickly buckles in the vertical direction. 
The snapshot at 1.8 Gyr is adopted to create the mock images. 
The scale length of the initial disk is $R_{\rm d, 0} = 1.9$ kpc. 
In Model B, the density distribution of the dark matter halo is described by an adiabatically compressed King profile ($\Phi(0) / \sigma^2 = 3$ and $r_{\rm t} = 10R_{\rm d}$, see \citealt{2005ApJ...634...70S} for details of adiabatic compression). The halo consists of 2.5 million particles, and its total mass $M_{\rm halo} = 8M_{\rm d}$.
The snapshot at 2.4 Gyr is adopted for this model.
Bars in both Model A and Model B have experienced buckling instabilities (\citealt{1991Natur.352..411R}).
The snapshots we choose have relatively strong spiral arms that are often seen in real observed galaxies.
The bar ellipticity ($\sim 0.5$ in Model A, $\sim 0.4$ in Model B) and the ratio of the bar length to the disk size ($\sim 0.5$ in Model A, $\sim 0.65$ in Model B) are also consistent with observations (Erwin 2005).
We estimated the ratio between the boxy/peanut and bar length of Model A ($\sim 0.57$) and Model B ($\sim 0.44$).
This ratio is consistent with other simulations (\citealt{2013MNRAS.431.3060E}).
Therefore, the two models are reasonably representative and adequate for the present study.
In addition, the bar in Model B is longer than that in Model A, which enables a consistency check of the different deprojection methods.

\figurenum{1}
\begin{figure}
\includegraphics[width=0.45\textwidth]{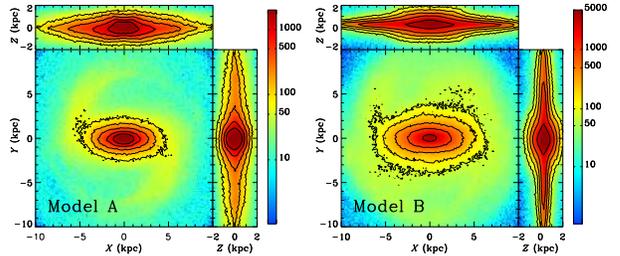}
\caption{Face-on and edge-on views of Model A (left) and Model B (right). Color represents the surface density.
}
\end{figure}

The mock images are created by projecting the 3D simulations onto a 2D plane ($200 \times 200$ pixels) with different disk inclination ($i$) and different bar orientations ($\phi_{\rm bar}$)\footnote{$\phi_{\rm bar}$ is introduced during the mock image creation. 
Initially, the bar model is aligned with the $X$-axis as shown in Figure~1. 
The model is rotated counterclockwise by $\phi_{\rm bar}$.
Then we incline it with respect to the $X$-axis (semi-major axis of the disk) by $i$ to create the mock image. 
After projection, the bar orientation ($\phi^\prime_{\rm bar}$) relative to the major axis of the inclined disk is slightly different from the initial $\phi_{\rm bar}$ due to the projection effect.
$\phi^\prime_{\rm bar}$ is measured directly from the mock images, and used to deproject the bar properties in our analysis. }.
The inclination angle $i$ varies from $0^\circ$ to $75^\circ$ and $\phi_{\rm bar}$ varies from $0^\circ$ to $90^\circ$. 
In the end, we generate 42 mock galaxy images with equally sampled $i$ and $\phi_{\rm bar}$ for each simulation. 

\subsection{Bar Measurement}

Bars leave distinct features in the isophotal geometric profiles. 
We fit isophotal ellipses of these mock images with IRAF task ELLIPSE. 
During the fitting, the center, position angle (PA) and ellipticity ($e$) of the isophotal ellipses are all set to be free parameters. 
In the literature, there are several methods commonly used to identify and measure a bar. 
Bars usually correspond to the maximum ellipticity ($e_{\rm max}$) and roughly constant position angle ($\Delta \rm PA \leq 10^\degree$) (\citealt{2005MNRAS.364..283E}; \citealt{2007ApJ...659.1176M}). 
We choose the position at $e_{\rm max}$ as a measurement of the bar length ($a_{\rm max}$), which tends to underestimate the visually identified bar length. 
Another method uses the position of the first ellipticity minimum ($e_{\rm min}$) outside $e_{\rm max}$ with roughly constant PA to probe a bar (\citealt{2005MNRAS.364..283E}). 
In this case, the bar length ($a_{\rm min}$) is set to be the position of $e_{\rm min}$. 
The transition from the bar to the disk generally corresponds to large variations in PA. 
Following \cite{2003ApJS..146..299E}, we measure another bar length ($a_{10}$) where PA varies by $10^\degree$ with respect to that within the barred region ($\rm PA_{\rm bar}$). 
The isophotal ellipticity at this radius is denoted as $e_{10}$.

Examples of our bar measurement using face-on images of the two models are shown in Figure~2. 
The vertical solid line marks $a_{\rm max}$, which requires $e_{\rm max} \geq 0.25$ and $\Delta \rm PA \leq 10^\degree$. 
The vertical dotted line denotes $a_{\rm min}$, corresponding to the first ellipticity minimum outside $e_{\rm max}$ with $\Delta \rm PA \leq 10^\degree$. 
$a_{10}$ is represented by the vertical dashed line, where $\Delta \rm PA = 10^\degree$. 
Thus, for each mock galaxy image, we have three different bar measurements based on ellipse fitting. 
The top right panel shows that $a_{\rm max}$ is well defined in Model A.
However, there is no clear peak (flat-topped) in the ellipticity profile of Model B in the bottom right panel.
Thus, we take $a_{\rm max}$ as the average value of the inner and outer radii where the ellipticity decreases to 90\% of the average value in the flat region.
It is our impression that the visual bar length is closest to $a_{\rm min}$. 
$a_{\rm max}$ is slightly shorter than the visual bar length, whereas $a_{\rm 10}$ tends to overestimate the visual bar length (almost twice as large as the visual value in Model A). 

\figurenum{2}
\begin{figure}
\includegraphics[width=0.45\textwidth]{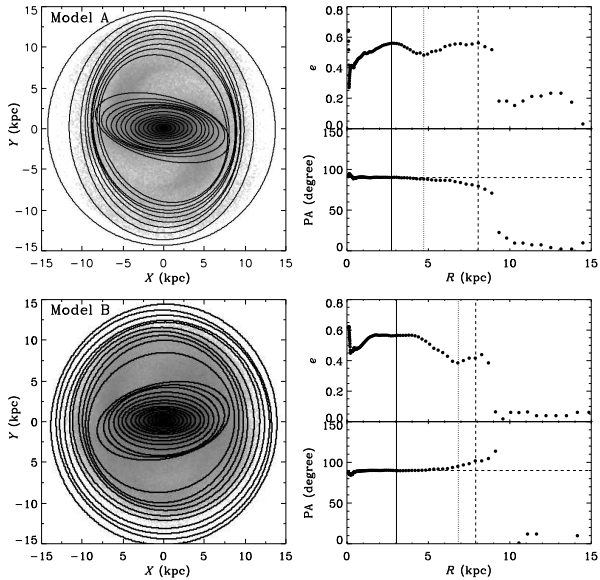}
\caption{Examples of the best-fit isophotal ellipses and radial profiles of $e$ and PA of each isophote for Model A (top row) and Model B (bottom row).
From these radial profiles we can determine three different bar lengths, which are $a_{\rm max}$ (vertical solid line), $a_{\rm min}$ (vertical dotted line) and $a_{\rm 10}$ (vertical dashed line).
The position angle of the bar is marked with the horizontal dashed line.
}
\end{figure}

In addition, we also use Fourier based methods to measure the bar length, namely the Fourier decomposition method and bar-interbar contrast method. 
Details about these two methods are described in Section \ref{sec:fourier}.

\section{BAR DEPROJECTION}

With our mock images, we test different deprojection methods in the literature. 
The basic assumption in these deprojection processes is that the outer part of a bar is assumed to be vertically thin. 
Bar properties of the inclined galaxies are measured with several different methods discussed in the previous section. 
To make a fair comparison, the same bar identification and measurement methods are also applied to the face-on images. 
Inclination angles of the disks are given as known quantities to avoid introducing additional uncertainties. 

\subsection{1D Analytical Deprojection}

This method analytically deprojects the measured major axis of the bar (\citealt{1995AJ....109.2428M}). The deprojected bar length is
\begin{eqnarray}
a_{\rm bar}^{\rm dep}=a_{\rm bar}^{\rm obs}(\cos^2\alpha+\sin^2\alpha \sec^2i)^{1/2},
\end{eqnarray}
where $a_{\rm bar}^{\rm obs}$ and $a_{\rm bar}^{\rm dep}$ are the observed and deprojected bar length, respectively. 
For galaxy images, $i$ is the disk inclination angle, and $\alpha$ is the angle between the projected major axes of the bar and the inclined disk ($\alpha = \phi^\prime_{\rm bar}$).

In Figure~3, we plot the ratios of the deprojected bar lengths ($a_{\rm max}^{\rm dep}$, $a_{\rm min}^{\rm dep}$, $a_{\rm 10}^{\rm dep}$) to the intrinsic face-on values ($a_{\rm max}^{\rm int}$, $a_{\rm min}^{\rm int}$, $a_{\rm 10}^{\rm int}$) as a function of $i$.
The left column illustrates the deprojection results of Model A.
Results of Model B are shown in the right column.
Please note that in certain extreme cases, e.g., large $i$ or $\phi_{\rm bar}$, it is difficult to measure the bar properties due to the complex ellipticity radial profiles.
Therefore we did not measure bar parameters for some cases in Figure~3.
As shown in panel (a) and (b), the deprojected $a_{\rm max}$ tends to overestimate the corresponding face-on value, with larger $i$ introducing higher uncertainties. 
The deprojection results at moderate inclinations ($i\leq60^\degree$) are generally overestimated by $\sim 40\%$. However, in an extreme case ($i = 60^\degree$, $\phi_{\rm bar} = 90^\degree$), the deprojected bar length can be overestimated by as much as 100\%. 
When $i > 60^\degree$, the deprojected $a_{\max}$ significantly overestimates the face-on values. 
In the case of $a_{\rm min}$, the general uncertainty is $\sim 25\%$ in moderately inclined disks ($i\leq 60^\degree$). 
Interestingly, at larger $i$, this uncertainty drops to $\sim 20\%$, which is much smaller than that of $a_{\rm max}$. 
For both $a_{\rm max}$ and $a_{\rm min}$, a larger $\phi_{\rm bar}$ usually results in a higher overestimation. 
The best case is $a_{\rm 10}$.  As shown in panel (e) and (f), the overestimation is quite small at moderate inclinations ($\sim 20\%$). 
The effect of $\phi_{\rm bar}$ is minimal in this case. 
Similar to $a_{\rm min}$, at large $i$ ($> 60^\degree$), the deprojection uncertainty of $a_{10}$ also decreases. 
The 1D analytical deprojection results of Model B show similar trend and scatter as in Model A.

\figurenum{3}
\begin{figure}
\includegraphics[height=0.45\textwidth,angle=-90]{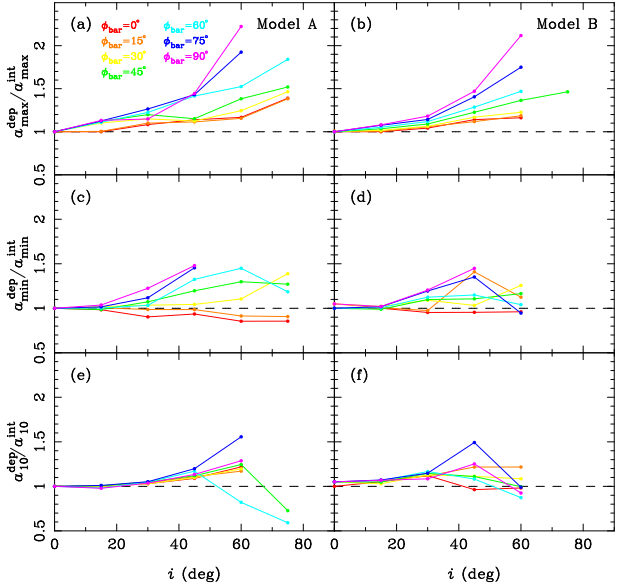}
\caption{Results of the 1D analytical deprojection of the bar length for Model A (left column) and Model B (right column).
From top to bottom, this figure shows the ratio of the deprojected bar length to the intrinsic face-on value ($a^{\rm dep} / a^{\rm int}$) as a function of the inclination angle for $a_{\rm max}$, $a_{\rm min}$ and $a_{10}$, respectively.
Different colors represent different $\phi_{\rm bar}$, which is the angle between the major axes of the bar and the inclined disk.
The black dashed line in each panel denotes unity, which means that the deprojection method perfectly recovers the true face-on bar length.
The missing data points at large $i$ ($\geq 70^\degree$) are due to the failure of the bar identification methods.
}
\end{figure}

\subsection{2D Deprojections}

\subsubsection{2D analytical deprojection}

The 2D analytical deprojection has been discussed in detail in \cite{2007MNRAS.381..943G}. 
We briefly review this method here. 
The shape of the bar is assumed to be a planar ellipse, which is analytically deprojected to an ellipse in the face-on view. 
The semi-major and semi-minor axes after the deprojection are:
\begin{eqnarray}
S1 = \left\{ \frac{2(\textit{A}{\textit{F}}^2+\textit{C}{\textit{D}}^2+\textit{G}{\textit{B}}^2-2\textit{B}\textit{D}\textit{F}-\textit{A}\textit{C}\textit{G})}{({\textit{B}}^2-\textit{A}\textit{C})[(\textit{C}-\textit{A})\sqrt{1+\frac{4{\textit{B}}^2}{(\textit{A}-{\textit{C})}^2}}-(\textit{C}+\textit{A})]} \right\}^{1/2}, \\
S2 = \left\{ \frac{2(\textit{A}{\textit{F}}^2+\textit{C}{\textit{D}}^2+\textit{G}{\textit{B}}^2-2\textit{B}\textit{D}\textit{F}-\textit{A}\textit{C}\textit{G})}{({\textit{B}}^2-\textit{A}\textit{C})[(\textit{A}-\textit{C})\sqrt{1+\frac{4{\textit{B}}^2}{(\textit{A}-{\textit{C}})^2}}-(\textit{C}+\textit{A})]} \right\}^{1/2},
\end{eqnarray}
where
\begin{eqnarray}
\textit{A} &= &\frac{\cos^2 \alpha}{\textit{a}^2} + \frac{\sin^2 \alpha}{\textit{b}^2}, \\
\textit{B}&= &\frac{\cos\alpha \sin\alpha \cos i}{\textit{a}^2} - \frac{\cos\alpha \sin\alpha \cos i}{\textit{b}^2}, \\
\textit{C}&= &\frac{\sin^2\alpha \cos^2i}{\textit{a}^2} + \frac{\cos^2\alpha \cos^2i}{\textit{b}^2}, \\
\textit{D}&= &\textit{F}=0, \\
\textit{G}&= &-1,
\end{eqnarray}
where $i$ is the inclination angle, and $\alpha$ represents the projected bar orientation ($\alpha = \phi^\prime_{\rm bar}$).
The semi-major and semi-minor axes lengths are $\max(S1, S2)$ and $\min(S1, S2)$, respectively. 
The deprojected ellipticity can then be calculated by $e = 1 - \min(S1, S2) / \max(S1, S2)$.

We perform the 2D analytical deprojection on all the mock images with different $i$ and $\phi_{\rm bar}$, and compare the deprojected values to the face-on ones.
Results shown in Figure~4 are much better than those in the 1D deprojection shown in Figure~3, because an ellipse simply describes the shape of the bar better than a straight line segment. 
For all the three different bar measurements, the agreement is quite good ($\sim$ 15\%) at small $i$ ($\leq 60^\degree$). 
At large $i$ ($>60^\degree$), the deprojection on $a_{\rm max}$ can overestimate the intrinsic face-on values by as much as 100\%. 
For $a_{\rm min}$, depending on $\phi_{\rm bar}$, the deprojection can either overestimate or underestimate the bar length by $\sim$ 10\% at moderate inclinations ($i\leq60^\degree$). 
$a_{\rm 10}$ tends to underestimate the bar length for Model A ($\sim$ 10\%), but to overestimate the bar length for Model B ($\sim$ 10\%). 
All the three deprojected results of the bar measurements depend weakly on $\phi_{\rm bar}$. 
Results of Model B are similar to Model A with slightly smaller scatters.

\figurenum{4}
\begin{figure}
\includegraphics[height=0.45\textwidth,angle=-90]{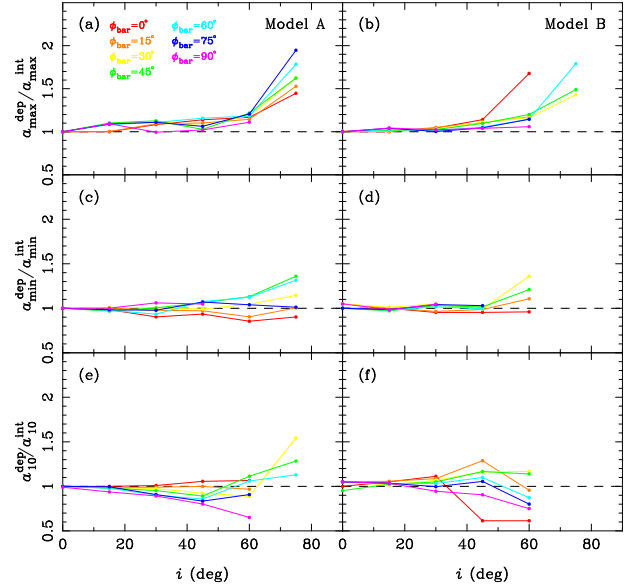}
\caption{As in Figure~3, but for the 2D analytical deprojection of the bar length.
}
\end{figure}

As shown in Figure~5, the deprojected ellipticity is accurate when $i \leq 60^\degree$. 
At moderate inclinations, the general deviation is less than 10\% for $e_{\rm max}$, $e_{\rm min}$ and $e_{\rm 10}$, but can be as large as 40\% in some cases. This method behaves badly at large $i$. 
The trend depends on $\phi_{\rm bar}$. 
Generally speaking, for $e_{\rm max}$ and $e_{\rm min}$, the method underestimates the face-on bar ellipticity for small $\phi_{\rm bar}$ ($\leq 50^\degree$), while it overestimates the ellipticity at large $\phi_{\rm bar}$ ($> 50^\degree$).
On the other hand, $e_{\rm 10}$ has much better agreement than $e_{\rm max}$ and $e_{\rm min}$. However, for Model B, at $i \sim 45^\degree$, $e_{10}$ shows very large uncertainties.
Since $e_{10}$ is the ellipticity at $a_{10}$, which is well beyond the visual bar length, the influence of the bar thickness to $e_{\rm 10}$ is much less than that to $e_{\rm max}$ and $e_{\rm min}$.
Comparing the left column with the right column, we can see that the deprojected ellipticities of Model A and Model B are very similar regardless of the different bar lengths.

\figurenum{5}
\begin{figure}
\includegraphics[height=0.45\textwidth,angle=-90]{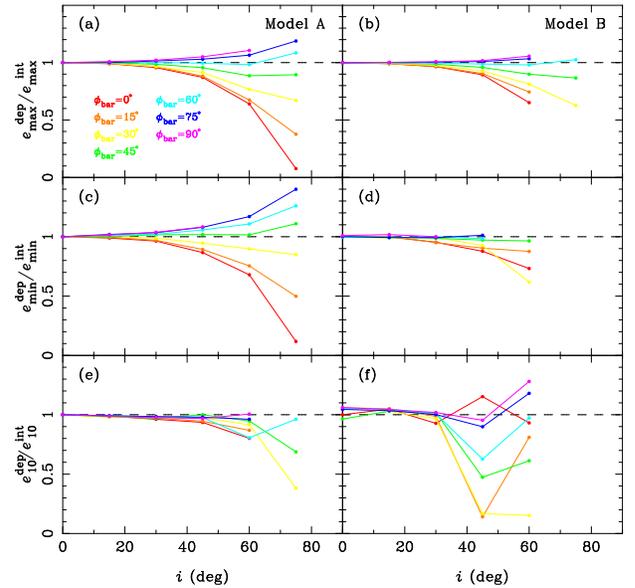}
\caption{As in Figure~3, but for the 2D analytical deprojection of the bar ellipticity.
}
\end{figure}

\subsubsection{2D image deprojection}

The mock inclined images can be first deprojected to the face-on images with the IRAF task GEOTRAN, which basically stretches the inclined image along the minor axis of the disk.
Then the bar properties are extracted from the stretched images.
The GEOTRAN routine enables us to correct the geometric distortion of galaxy images while keeping the total flux conserved.
Firstly, the major axis of the inclined galaxy is rotated to the direction of $X$-axis.
Then the size of the minor axis of the disk is linearly magnified to the original value according to the disk inclination.
Hence, this method deprojects the whole galaxy image to face-on directly (see Figure~6 for an example). 
Another image deprojection method is the IRAF task IMLINTRAN (\citealt{2007MNRAS.381..943G}).
We have also tested this method and found that the deprojected images using these two routines are almost identical.
The former task is adopted in this work.
Comparing the deprojected bar properties extracted from these stretched images to the ones measured in the original face-on images, we can estimate the uncertainty of this method. 
However, once $i$ exceeds $60^\degree$, the images after stretching completely betray the real face-on ones. 
In such cases, it is almost impossible to identify the bar with the measured radial profiles of ellipticity and PA. 
Thus, the 2D image deprojection is confined to small $i$ only ($\leq 60^\degree$). 

\figurenum{6}
\begin{figure}
\includegraphics[width=0.45\textwidth]{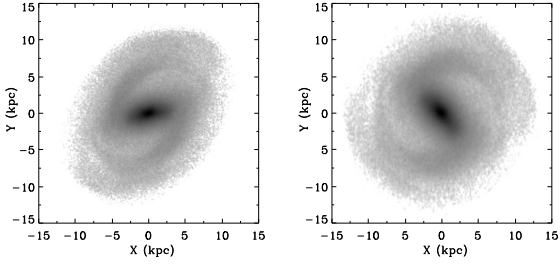}
\caption{An example of the 2D image deprojection.
The left-hand panel shows the inclined image of Model A ($i = 45^\degree$, $\phi_{\rm bar} = 45^\degree$).
The right-hand panel shows the image being stretched to the face-on image by IRAF task GEOTRAN.}
\end{figure}

Figure~7 and 8 show the results of the deprojected bar length ($a_{\rm max}$, $a_{\rm min}$ and $a_{10}$) and ellipticity ($e_{\rm max}$, $e_{\min}$ and $e_{10}$), respectively.
For Model A, the left column of Figure~7 shows that the deprojected $a_{\rm max}$, $a_{\rm min}$ and $a_{10}$ agree with the face-on values very well; the uncertainty is about 10\% at moderate inclinations. 
Systematic overestimation is found in $a_{\rm max}$, while for $a_{\rm min}$ and $a_{10}$, the situation is uncertain. 
For Model B, the right column indicates that the results have similar trends but different scatters compared to Model A.
The deprojected results of $a_{\rm max}$ and $a_{\rm min}$ are very good ($\sim 5 \%$) except that the deprojected $a_{\rm 10}$ has a relatively large uncertainty (up to $\sim$ 10\%) at moderate inclinations.
Generally speaking, the deprojection of Model B is more accurate than Model A.
All the panels show that the deprojected bar length depends weakly on $\phi_{\rm bar}$.

\figurenum{7}
\begin{figure}
\includegraphics[height=0.45\textwidth,angle=-90]{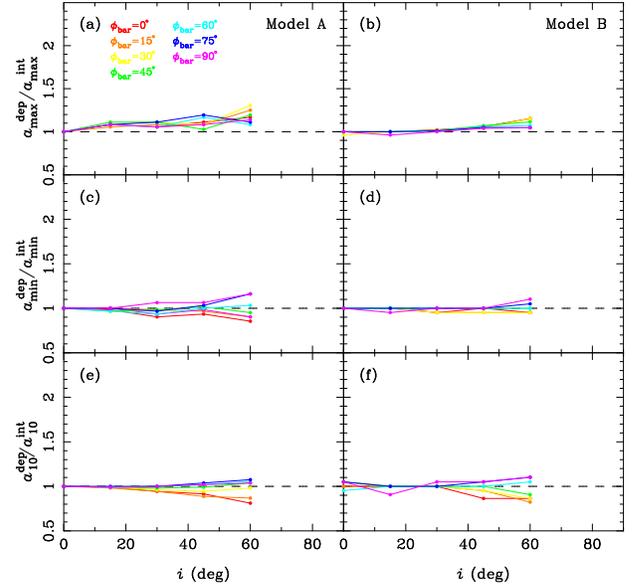}
\caption{As in Figure~3, but for the 2D image deprojection (image stretching) of the bar length.
}
\end{figure}

\figurenum{8}
\begin{figure}
\includegraphics[height=0.45\textwidth,angle=-90]{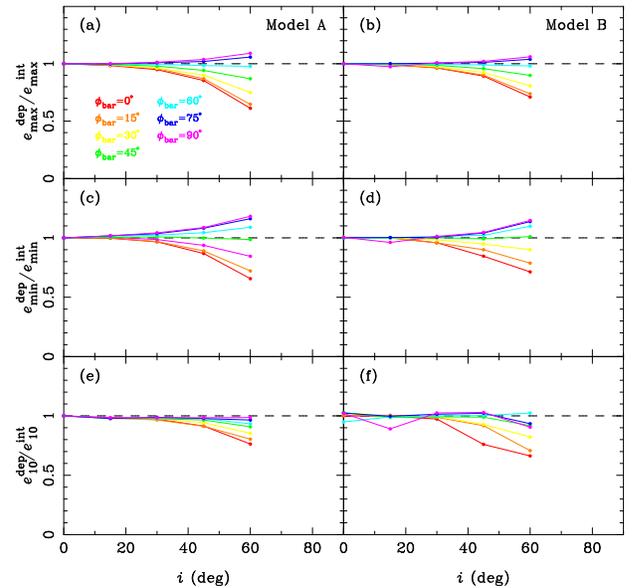}
\caption{As in Figure~3, but for the 2D image deprojection (image stretching) of the bar ellipticity.
}
\end{figure}

For $e_{\rm max}$ and $e_{\rm min}$, the 2D image deprojection tends to underestimate the face-on values by $\sim 10\%$ at small $\phi_{\rm bar}$ ($\leq 50^\degree$).
This trend reverses at large $\phi_{\rm bar}$ ($> 50^\degree$).
The deprojected $e_{\rm 10}$ seems to underestimate the face-on value by $\sim$ 10\% regardless of $\phi_{\rm bar}$.
As shown in Figure~8, the deprojected ellipticity behaves similarly to the 2D analytical deprojection in Figure~5. 
This figure suggests that the deprojected ellipticity is more accurate when $i$ is less than $60^\degree$. 
There is no difference in the deprojection of the ellipticity between Model B and Model A.

\subsection{Fourier Based Deprojections}
\label{sec:fourier}

\subsubsection{Fourier decomposition}

In this work, we also test the Fourier decomposition method in recovering the bar length measured in the inclined image (\citealt{2007MNRAS.376.1480N}; \citealt{2011ApJS..197...22L}). 
We fit the galaxy images with the center, PA and $e$ of each elliptical annulus fixed to the values measured at the outskirts of the disk. 
Then we decompose the intensities within each elliptical annulus with the equation
\begin{eqnarray}
\textit{I}(\theta)=\textit{I}_{0}+\sum\limits_{m=1}^{\infty} \textit{I}_{m}\cos\left(m\theta+\phi_{m}\right) ,
\end{eqnarray}
where $I$ is the intensity on the annulus in the direction of $\theta$. 
\textit{I}$_{0}$ is the averaged intensity of each annulus. 
\textit{I}$_{\rm m}$ is the amplitude of $m$-th mode of the Fourier series. 
$\phi_{\rm m}$ is the corresponding phase angle. 
Figure~9 shows an example of the Fourier decomposition. 
The upper left hand panel illustrates the image with the fixed ellipses overlaid. 
The upper right hand panel is the radial profile of the relative amplitude of the Fourier $m = 2$ mode. 
The bar corresponds to large $I_2/I_0$. 
The peak position of $I_2/I_0$ is within the bar region, where the bar-interbar contrast is the strongest. 
Based on our empirical tests, we choose the position at $0.85(I_2/I_0)_{\rm max}$ outside the peak position as the end of the bar. 
Assuming the disk is purely circular in its face-on view, the semi-major axes of the ellipses (bottom left panel) in fact equals to the radii of the face-on circular annulus (bottom right panel). 
The bar length is marked by the semi-major axis of the particular ellipse that encloses the bar region.
This value is actually the radius of the circle passing right through the bar ends in the face-on view. 
Thus the bar length measured in the inclined image is the same as the length in the face-on image, which can be directly compared to the face-on values.

\figurenum{9}
\begin{figure}
\includegraphics[width=0.45\textwidth]{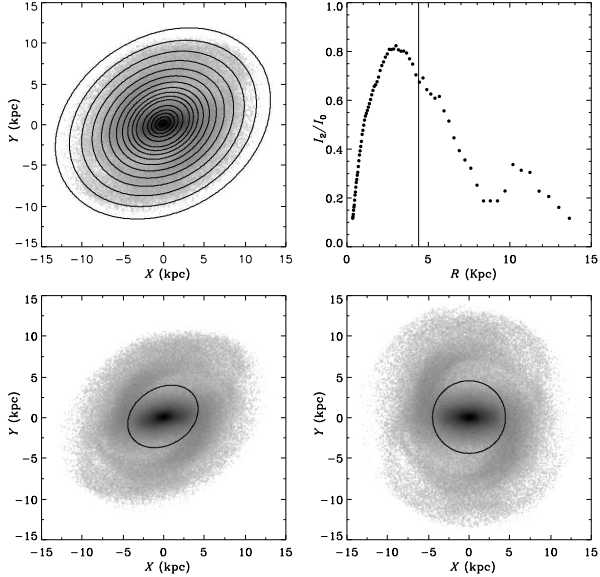}
\caption{An example of the isophote measurement with the geometric parameters fixed to the outer most isophote (upper left panel) and the radial profile of the relative amplitude of Fourier $m=2$ mode (upper right panel) for Model A ($i=45^\degree$, $\phi_{\rm bar}=30^\degree$).
The bar end marked by the solid line corresponds to 0.85(I2/I0)max.
The bottom right panel shows the face-on image of Model A, overlaid with a circle enclosing the bar ends.
The bottom left panel shows the projection of the image in the bottom right panel.
}
\end{figure}

As shown in Figure~10, the bar length measured in the inclined image agrees quite well with that of the face-on image. 
The typical difference at moderate inclination is $\sim$ 10\%. 
However, when $i$ is larger than $60^\degree$, the measured bar length overestimates the face-on value by as much as 50\%. 
Similar to the previous results, at lower $i$ ($\leq 60^\degree$), the influence of $\phi_{\rm bar}$ is negligible. 
Most of the uncertainties in the deprojection come from the large inclination angles. 

\figurenum{10}
\begin{figure}
\includegraphics[height=0.45\textwidth,angle=-90]{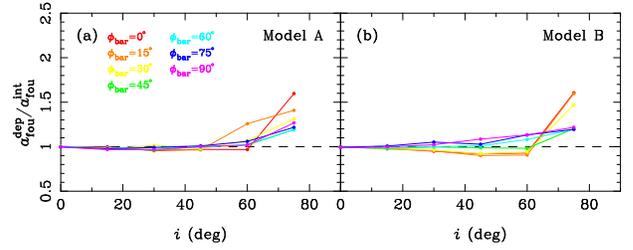}
\caption{Results of the Fourier decomposition method.
This figure shows the deprojected bar length ($a^{\rm dep}_{\rm fou}$) to the intrinsic face-on value ($a^{\rm int}_{\rm fou}$) as a function of the inclination angle (\textit{i}).
The left-hand panel shows the result of Model A, while the right-hand panel shows the result of Model B.
Different colors represent different $\phi_{\rm bar}$, which is given in panel (a).
The black dashed line denotes unity.
}
\end{figure}

\subsubsection{Bar-interbar contrast}

In the Fourier decomposition method, the largest contribution to the bar comes from the Fourier $m = 2$ component. 
As demonstrated in Figure~11, the higher even-order components have weaker amplitudes. 
\cite{1990ApJ...357...71O} argued that the density distribution of the bar should be approximated by all the important even Fourier components. 
Based on these even Fourier modes, they further suggested the construction of radial profiles of the luminosity contrast between the bar and interbar regions. 
The bar intensity ($I_{\rm b}$) is defined as $I_0 + I_2 + I_4 + I_6$, and the interbar intensity ($I_{\rm ib}$) is $I_0 - I_2 + I_4 - I_6$. 
They defined the bar region as the zone where $I_{\rm b} / I_{\rm ib}$ is larger than 2. 
However, setting the value as 2 is not physically meaningful because it can not account for all the morphological differences among galaxies.
\cite{2000A&A...361..841A} proposed a more reasonable method to find the bar region:
\begin{eqnarray}
I_{\rm b} / I_{\rm ib} > \frac{(I_{\rm b} / I_{\rm ib})_{\rm max}-(I_{\rm b} / I_{\rm ib})_{\rm min}}{2}+(I_{\rm b}  / I_{\rm ib})_{\rm min},
\end{eqnarray}
which is equivalent to the full width at half-maximum (FWHM) of $I_{\rm b} / I_{\rm ib}$ profile.

\figurenum{11}
\begin{figure}
\centerline{\includegraphics[width=90mm,angle=-90]{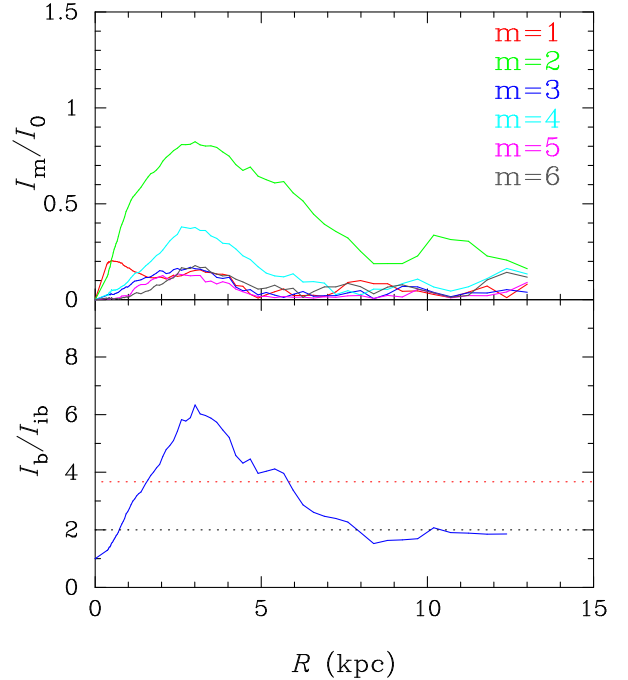}}
\caption{An example of the bar-interbar contrast for the same image as Figure~9.
The top panel shows the radial profile of the relative amplitude of the first six modes of the Fourier series.
The bottom panel shows the radial profile of the luminosity contrast between the bar and the inter-bar region.
The black dotted line represents the criteria in the literature.
The red dotted line is the value determined by Equation (10). }
\end{figure}

The left-hand panel of Figure~12 shows that, overall, this method tends to overestimate the deprojected bar length of Model A. 
The amount of the overestimation is comparable to that of the 2D analytical deprojection. 
When $i$ is larger than $60^\degree$, the deprojected bar length is typically overestimated by about 15\%. 
This panel also shows different behavior compared to the previous results on $\phi_{\rm bar}$. 
The overestimation actually decreases with increasing $\phi_{\rm bar}$. 
But it plays a minor role in determining the deprojected bar length.
The right-hand panel of Figure~12 indicates that the deprojected bar length of Model B is more accurate than that of Model A.
At small $i$ ($\leq 60^\degree$), the deprojected lengths agree very well with the face-on value.
When $i > 60^\degree$, similar to Model A, the deprojected lengths are overestimated by as much as 50\%.

\figurenum{12}
\begin{figure}
\includegraphics[height=0.45\textwidth,angle=-90]{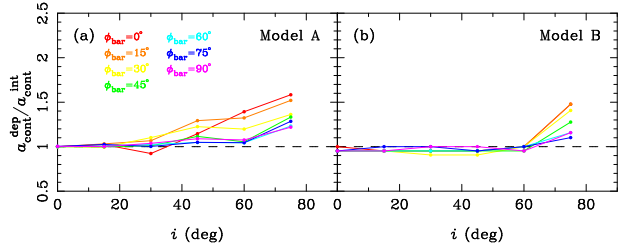}
\caption{As in Figure~10, but for the bar-interbar contrast method.
}
\end{figure}

\section{DISCUSSION}

\subsection{Deprojection Uncertainties}

In this work, we investigate uncertainties in the deprojection of the bar properties of two models, i.e., Model A and Model B.
Figure~13 summarizes the typical scatter ranges in different deprojection methods for the bar length and ellipticity at moderate inclinations ($i\leq60^\degree$).
The scatter comes from measures corresponding to different values of $\phi^\prime_{bar}$ and $i$.
For each method, the upper (lower) limit of the scatter is the first (third) quartile of the overall distribution of absolute values of scatters at $30^\degree \leq i \leq 60^\degree$.
This figure shows that the 1D analytical deprojection always has the largest scatters in the two models, which is independent on the length of the bar. 
The 1D deprojection has the simplest assumption of the bar structure; the only measured property of the bar used in this deprojection process is the major axis (Equation 1), which alone does not reflect the 3D structure of the bar at all. 
In addition, the 2D analytical deprojection and the 2D image deprojection produce consistent results for the two models. 
The scatters of these two methods are relatively small at $i\leq60^\degree$.
However, the 2D image deprojection (image stretching) is almost impossible to deproject the bar properties at $i>60^\degree$. 
All the barred galaxies in our simulations show a vertical X-shaped structure in the inner region when $i>60^\degree$. 
Thus, the stretched image is very different from the true face-on image. 
Figure~14 shows an example of the 2D image deprojection at $i=75^\degree$.
Comparing the top-left panel to the bottom-left panel, we can see that the deprojected image is completely different from the original face-on one.
The bottom-right panel also shows different features in the ellipticity profile, in which the bar is difficult to recognize.

\figurenum{13}
\begin{figure}
\includegraphics[width=0.45\textwidth]{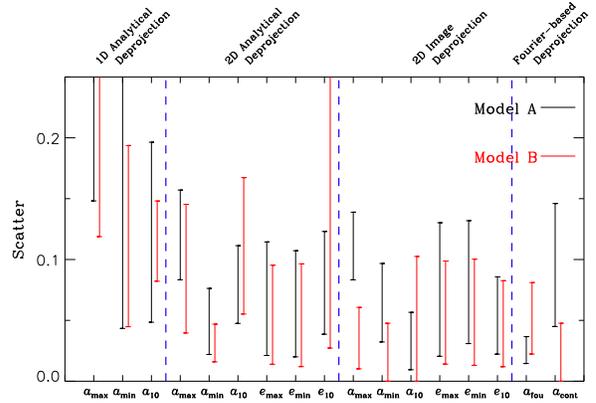}
\caption{Typical scatters in different deprojection methods for both bar length and ellipticity at $i \leq 60^\degree$.
The black and red colors represent Model A and Model B, respectively.
Error bars show the scatter range.
The upper (lower) limit represents the first (third) quartile of the distribution of all the scatter values for a given method at $i \leq 60^\degree$.
Please notice that the scatter of each method comes from measures corresponding to different values of $\phi^\prime_{bar}$ and $i$.}
\end{figure}

\figurenum{14}
\begin{figure}
\includegraphics[height=0.45\textwidth]{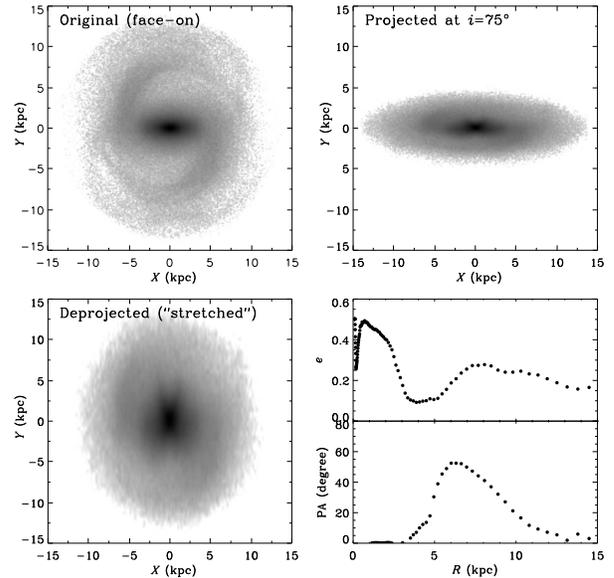}
\caption{Examples of the 2D image deprojection at $i=75^\degree$.
The top-left panel shows the original face-on image of Model A.
The top-right panel shows the image of Model A at $i=75^\degree$ and $\phi_{\rm bar}=0^\degree$.
The deprojected image is shown in the bottom-left panel and the corresponding geometric radial profiles of the best-fit isophotal ellipses are shown in the bottom-right panel.}
\end{figure}

The deprojection results shown in Section 3 suggest that scatters of the Fourier decomposition is $\sim$ 10\% at moderate inclinations ($i\leq 60^\degree$), which is generally consistent between the two models.
Another Fourier-based method, namely the bar-interbar contrast method, has relatively larger uncertainties than the Fourier decomposition method.
On the other hand, uncertainties of the bar-interbar contrast method seems to be model-dependent, because this method considers the even ($m=2, 4, 6$) modes to calculate the bar-interbar contrast, which in fact depends on the detailed stellar distribution of the bar.

For the deprojected ellipticity, the 2D analytical deprojection and the 2D image deprojection produce very similar results.
Uncertainties increase at large $i$.
As $\phi_{\rm bar}$ increases from $0^\degree$ to $90^\degree$, the deprojected ellipticity gradually transitions from underestimation to overestimation. 
In addition, at certain $i$, uncertainties of the deprojected bar ellipticity are generally larger than the deprojected bar length, because the vertical structure of the bar has more influence on the measured isophotal ellipticity than the length.

Uncertainties in the deprojected ellipticities of these two models are almost the same.
This confirms the trend in the ellipticity deprojection as $\phi_{\rm bar}$ increases from $0^\degree$ to $90^\degree$.
This is not surprising because at large $\phi_{\rm bar}$, the bar orientation aligns close to the disk minor axis. 
Therefore, the thickness of the bar increases the projected bar length when the galaxy is inclined, which makes the measured ellipticity larger than that expected from a planar bar.
The results are reversed at small $\phi_{\rm bar}$ since the thickness of the bar enlarges the bar minor axis length when the galaxy is inclined.
Thus the measured ellipticity becomes smaller.

In general, most of the results presented in this work suggest that the deprojection uncertainties of Model A and Model B have similar scatters and trends.
Despite this consistency, scatters in galaxies with a long bar (Model B) are slightly smaller than that with a short bar (Model A), which could be due to the fact that the outer part of the bar in Model B is less affected by the inner thickened bulge than in Model A.

Theoretically, one of the deprojection uncertainty stems from the calculation of the inclination angle of the outer disk. 
The simplest way assumes that the disk is round and thin.
Then, the inclination angle can be derived from a simple formula, i.e., ${\rm cos}(i)=1-e_{\rm disk}$.
This method is applied to late type galaxies because their disk is relatively thin.
The \cite{1926ApJ....64..321H} method utilizes the ellipticity of the disk outskirts to derive the inclination angle under the assumption of a certain intrinsic thickness and shape of the disk, which gives a more accurate estimation of the inclination angle for early type galaxies.
Since the intrinsic thickness and roundness of the disk are unknown, these methods inevitably introduce uncertainties to the bar deprojection. 
To test these uncertainties, the differences between the observed $i$ and the corresponding given values are investigated in this work.
The first method mentioned above is adopted to measure the inclination angle.
We find that the inclination uncertainty is very small ($\sim 5^\degree$) at intermediate inclinations ($i\leq75^\degree$).
However, at very small inclination angles, the difference is relatively large due to the simple assumption.
In our simulation, the outer skirt of the galaxy are not perfect featureless and circular (e $\sim 0.15$).
A slight distortion of the isophotes in the outer part will result in a relatively large inclination angle ($\sim 20^\degree$).
However, the influence of $i$ on the deprojection should be trivial. 
In our work, $10^\degree$ difference in inclinations does not affect the deprojected bar parameters too much. 
We assume that the inclination angle is exactly known.
Even though the influence of the inclination is limited in our models, it is worth pointing out that the uncertainty of deprojection derived in this work is only a lower limit.
The true uncertainty will be higher if the inclination error is considered.
The error in $\phi^\prime_{bar}$ measurement is also studied.
The difference between the measured bar angle and the given value during the mock image creation is very small ($\sim 5^\degree$) at intermediate inclination ($i<75^\circ$), but it becomes relatively large at very small $i$.
When the disk is close to face-on, the line of nodes (LON) of disk is highly uncertain.
A slight change in the outskirt shape could result in huge difference between the bar angle and the disk LON. 
In this work, $\phi^\prime_{bar}$ is measured from the mock images.
The error in $\phi^\prime_{bar}$ measurement will not influence our deprojection results.

Generally speaking, the measurement error in $i$ and $\phi^\prime_{bar}$ could introduce uncertainties to the deprojection.
However, such uncertainties are small compared to the one caused by 3D structure of the bar itself.
We will carefully investigate this with toy models in the next section.

\subsection{Toy Models of the 3D Bar Structure}
Simulations found that an evolved bar is thick in the inner part due to the vertical buckling instability (\citealt{1981A&A....96..164C}; \citealt{1991Natur.352..411R}; \citealt{2004ApJ...605..714D}; \citealt{2005MNRAS.358.1477A}; \citealt{2006ApJ...645..209D}). 
Observations of intermediately inclined barred galaxies also found non-negligible thickness of the bar (\citealt{1999AJ....118..126B}; \citealt{2006MNRAS.370..753B}). 
We want to know to what extent this vertical structure influences the accuracy of the bar deprojection.

\begin{deluxetable}{cccccccc}
\tablecolumns{10}
\tablewidth{0pc}
\tablecaption{Geometric parameters of the toy models}
\tablehead{
\colhead{} &
\colhead{$a_{\rm max}$} &
\colhead{$b_{\rm max}$} &
\colhead{$a_{\rm min}$} &
\colhead{$b_{\rm min}$} &
\colhead{$a_{\rm 10}$} &
\colhead{$b_{\rm 10}$} &
\colhead{$h$} \nl
\colhead{} &
\colhead{(kpc)} &
\colhead{(kpc)} &
\colhead{(kpc)} &
\colhead{(kpc)} &
\colhead{(kpc)} &
\colhead{(kpc)} &
\colhead{(kpc)} \nl
\colhead{} &
\colhead{(1)} &
\colhead{(2)} &
\colhead{(3)} &
\colhead{(4)} &
\colhead{(5)} &
\colhead{(6)} &
\colhead{(7)}
}
\startdata
Model A& 2.7& 1.2& 4.7& 2.4& 8.0& 3.5& 1.5\\
Model B& 3.0& 1.4& 6.8& 4.3& 7.9& 4.6& 2.0
\enddata

\tablecomments{Cols. (1) -- (2): Major and minor axis of the bar traced by maximum ellipticity. Cols. (3) -- (4): Major and minor axis of the bar traced by minimum ellipticity. Cols. (5) -- (6): Major and minor axis of the bar traced by $10^\degree$ position angle variation. Col. (7): Vertical thickness of the bar.
}
\end{deluxetable}

In the previous section, we conclude that all the deprojection methods behave badly at large $i$ ($>60^\degree$). 
To better understand this result, we first look at the projected bar at different viewing angles ($i$ and $\phi_{\rm bar}$). 
In the edge-on image, a bar structure contains a boxy bulge in the inner region. 
To simplify the calculation, we use toy models here to show the projection process. 
Figure~15 shows the sketch of a toy model.
There are two different parameter sets corresponding to Model A and Model B.
The values are listed in Table 1.
The structure of the bar is treated as a triaxial ellipsoidal shell with the axis ratio as $a:b:h$, where $a$ is the semi-major axis, $b$ is the semi-minor axis and $h$ is the vertical thickness.
The values in Table 1 are estimated from the measured bar properties in the face-on and edge-on mock images.
Then the triaxial ellipsoids with axis ratio of $a:b:h$ are constructed.
Several simulations suggest that the outer part of the bar should be much thinner than the inner region (\citealt{2005MNRAS.358.1477A}).
However, we found that one-ellipsoid models could represent the 3D structure of the  bar reasonably well.\footnote{We also tested two ellipsoid in the toy model construction, i.e., an inner thick one and an extended thin one.
The deprojection results are consistent with the single ellipsoid toy model.}
At large inclination angles, the thickness of the outer bar region plays a minor role in deprojection uncertainties; it is mainly the inner thick part that which significantly changes the projected bar shape.
At small inclinations, the effects of outer bar thickness in the deprojection uncertainties are also limited.

\figurenum{15}
\begin{figure}
\centerline{\includegraphics[width=0.45\textwidth]{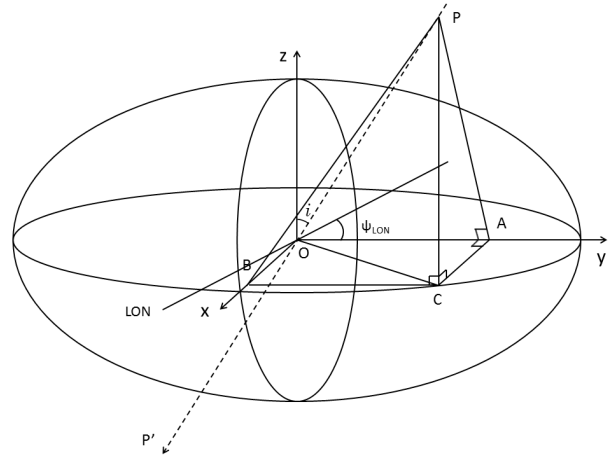}}
\caption{The sketch of the toy model.
The major axis, minor axis and the height are along $Y$-, $X$- and $Z$-axis, respectively.
The dashed line shows the viewing direction (PP$^\prime$), which passes through the center O.
Segment OC is the projection of OP onto the $X-Y$ plane.
}
\end{figure}

After projecting these 3D ellipsoids to a 2D plane from different $i$ and $\phi_{\rm bar}$, the major axis and the minor axis of the projected 2D ellipses can be calculated.
We deproject these 2D bars to the face-on properties using both the 1D and 2D analytical deprojection methods and compare them to the true face-on values.
The 1D analytical deprojection results of the toy models are shown in Figure~16. The deprojection trend and scatter of the toy models are very similar to the results of our simulations shown in Figure 3.

\figurenum{16}
\begin{figure}
\includegraphics[height=0.45\textwidth,origin=c,angle=-90]{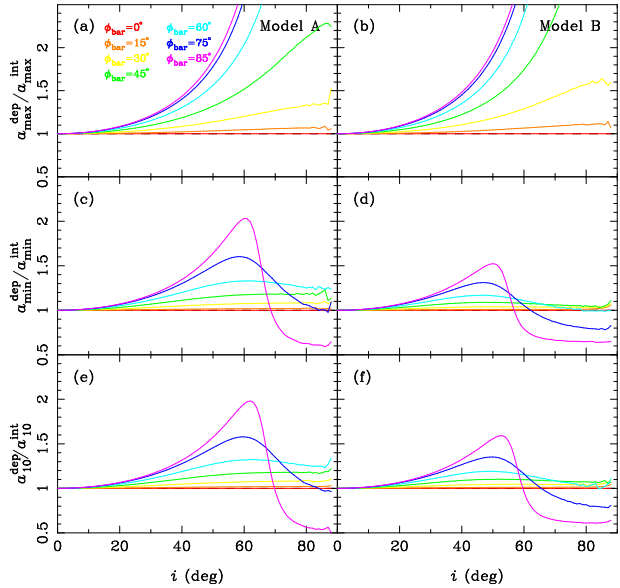}
\caption{As in Figure~3, but for the 1D analytical deprojection of the bar length in the toy models.
}
\end{figure}

It is obvious that the true shape of a bar is not a simple straight line segment; it has some finite width and height, making the semi-major axis and orientation of the face-on bar differ from those of inclined bar\footnote{At extremely large $i$ and $\phi_{\rm bar}$, the projected major axis can be even smaller than the projected minor axis, which will make us mistakenly treat the projected minor axis as the intrinsic major axis. In our toy models, we deproject the bar length at $\phi_{\rm bar}=85^\degree$ rather than $\phi_{\rm bar}=90^\degree$.}. 
In addition, the toy models do not consider the spatial density variations inside the bar and the projection effect from the surrounding disk, which could add additional errors to the bar measurement.
These will be discussed in the next section.

The 2D analytical deprojection is also tested using the toy models.
Comparing to Figure~16, we can see that the 2D analytical deprojection is indeed better than 1D as shown in Figure~17.
This figure shows that the 2D analytical deprojection tends to give higher overestimation on the bar length at larger $\phi_{\rm bar}$.
At small $i$, the scatter of the deprojection is relatively small.
However, this deprojection method is unable to recover the bar length at large $i$.
Comparing to our previous simulations in the deprojected bar length, it is nice to see that Figure~17 generally matches Figure~4. On the other hand, Figure~18 shows the 2D analytical deprojection results of the ellipticity based on the toy models. Comparing to Figure~5, we also find good agreements between deprojected ellipticities of the toy models and our simulations.

\figurenum{17}
\begin{figure}
\includegraphics[height=0.45\textwidth,origin=c,angle=-90]{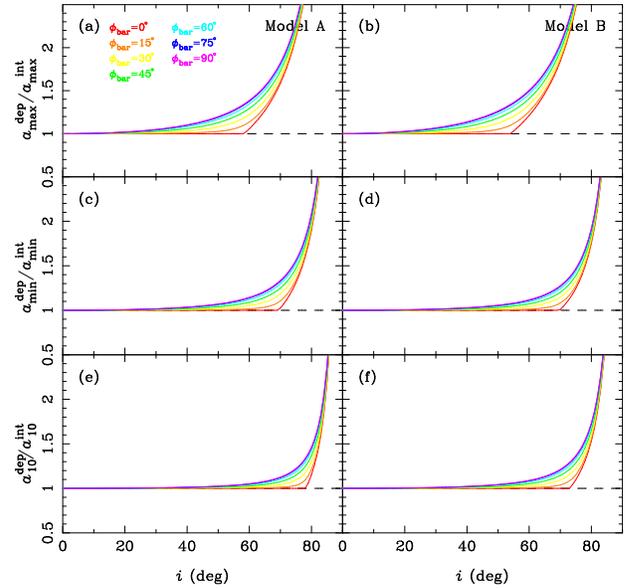}
\caption{As in Figure~3, but for the 2D analytical deprojection of the bar length in the toy models.
}
\end{figure}

\figurenum{18}
\begin{figure}
\includegraphics[height=0.45\textwidth,origin=c,angle=-90]{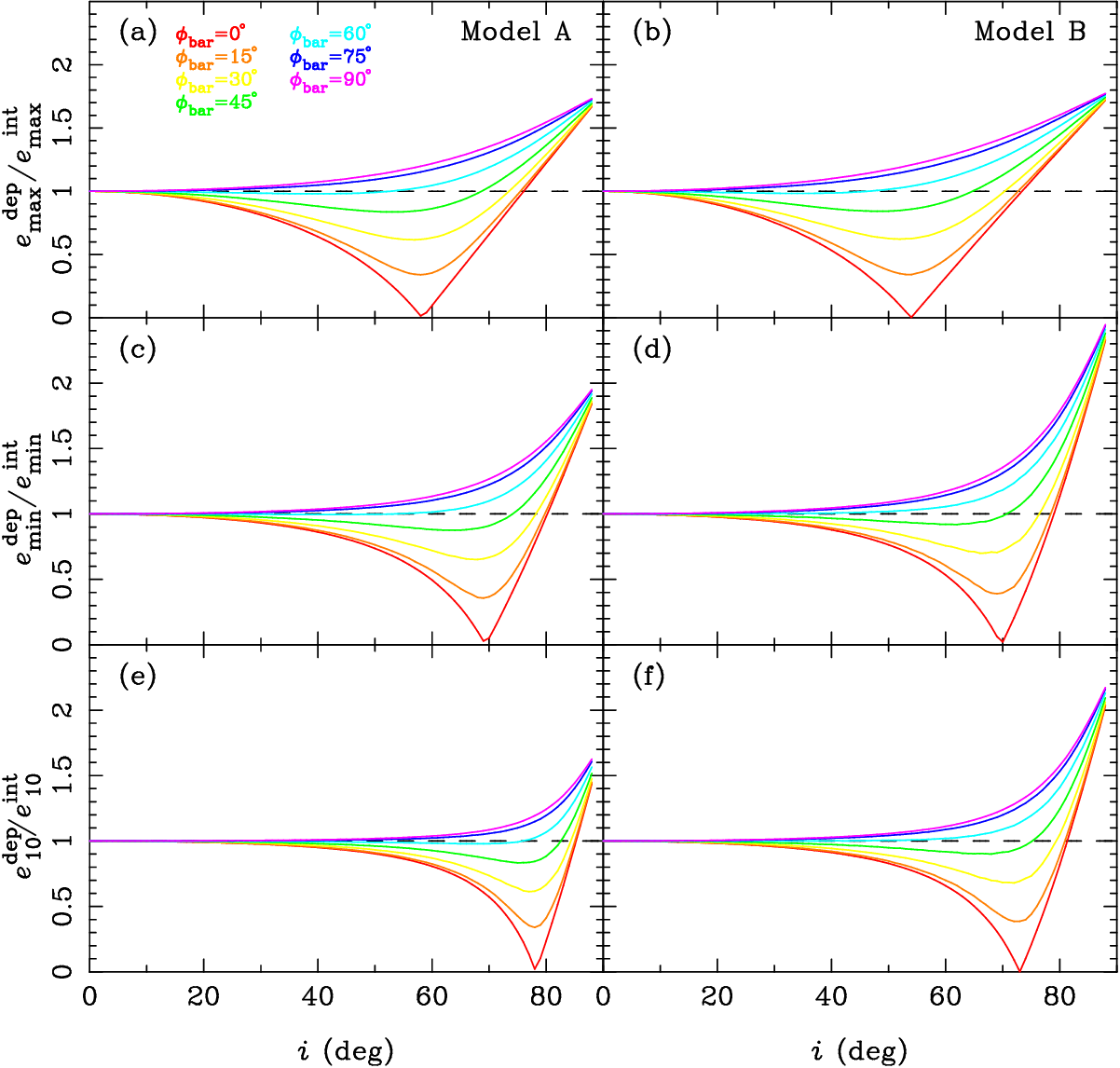}
\caption{As in Figure~3, but for the 2D analytical deprojection of the bar ellipticity in the toy models.
}
\end{figure}

Additionally, comparison among the results of $a_{\rm max}$, $a_{\rm min}$ and $a_{\rm 10}$ suggests that the turning point at which the deprojection uncertainties become large is different for these three kinds of bar length.
The deprojected $a_{\rm max}$, $a_{\rm min}$ and $a_{\rm 10}$ have turning points at about $50^\degree$, $60^\degree$ and $70^\degree$ inclination angles, respectively.
As shown in Figure~2, $a_{\rm max}$ is the shortest (inside the visually identified bar), which can be easily affected by the thick part of the bar.
$a_{\rm 10}$ is the longest (outside the visually identified bar).
Hence the influence by the thickness of the bar is the smallest.
This suggests that the 2D analytical deprojection can be affected by the identification of the bar.
Results of Figure~4 show that, the deprojected $a_{\rm max}$ tends to overestimate the true face-on $a_{\rm max}$, while the deprojected $a_{\rm 10}$ is prone to underestimate the true face-on $a_{\rm 10}$.
However, results of our toy models do not show such trends.
This is probably related to uncertainties in the bar measurement.
Briefly speaking, in the case of $a_{min}$ and $a_{10}$, the deprojected results underestimate the true face-on value at low $\phi_{bar}$.
That is because the bar lengths ($a_{min}$ or $a_{10}$) directly measured in the inclined images shrink with respect to the true face-on bar length after projection.
Thus, it is reasonable that the results of our toy models show some different features when compared to the simulated galaxies.
The uncertainties of bar measurement will be discussed in the next section.

Since the distribution of stars within the bar is too complicated for a toy model to represent, we did not directly test the 2D image deprojection and other Fourier-based deprojection methods with our toy models.
Fourier decomposition analyzes the azimuthal variations in the light distribution and compares the intensity between the bar and inter-bar region.
It gives us more information about the material distribution in the bar component (Fourier $m=2$ mode).
Near the ends of the bar, Fourier $m=2$ mode reaches its maximum value.
Thus, in principle, the influence of the bar thickness should be less important on the Fourier based methods at small $i$ ($\leq60^\degree$).

\subsection{Bar Measurement Uncertainties}
Except the uncertainties mentioned above, there are some uncertainties in the bar measurement which could also influence the accuracy of deprojection.
Firstly, the methods used here produce bar parameters with noticeable differences. 
$a_{\rm min}$ seems to represent the visual bar length well (\citealt{2005MNRAS.364..283E}). 
$a_{\rm max}$ is located inside the bar region, which tends to underestimate the visual bar length (\citealt{1995A&AS..111..115W}; \citealt{2002MNRAS.330...35A}). 
$a_{10}$ is always found in the disk region outside the bar, which tends to overestimate the visual bar length. 
Thus the accuracy of the deprojection also relies on the choice of the bar identification method. 

From our results of the 2D analytical deprojection, the deprojected $a_{\rm max}$ generally over-predicts the true face-on $a_{\rm max}$. 
$a_{\rm min}$ and $a_{10}$ can either over- or underestimate the true face-on values, depending on $\phi_{\rm bar}$. 
After excluding the disk particles outside the barred region in our model, we find that $a_{\rm max}$ measured from particles in the bar is located in the boxy bulge region, where the bar is thickened in the vertical direction. 
Thus, the uncertainty of the deprojected $a_{\rm max}$ is mainly affected by the 3D structure of the bar itself.
The deprojection results of $a_{\rm max}$ in the toy model gives the best agreement with our simulation.
$a_{\rm min}$ is quite close to the visually identified bar end.
At small $\phi_{bar}$, the measured $a_{min}$ in the inclined images is affected by the interplay between the bar and the disk, which is usually smaller than the true face-on $a_{min}$ after projection.
Therefore, the deprojected $a_{\rm min}$ at small $\phi_{bar}$ could underestimate the true face-on value.
Apparently, the errors in the measure of $a_{\rm 10}$ originate from the disk since $a_{\rm 10}$ is always larger than $a_{\rm min}$. 
$a_{10}$ measured from the inclined images are also smaller than the directly deprojected true face-on $a_{10}$ after projection.
That's the reason for the underestimation of deprojected $a_{10}$ at low $\phi_{bar}$.

For the Fourier decomposition, our results show that this method also produces consistent results.
The main reason is due to the bar measurement method, where the bar length is determined by the relative amplitude of the Fourier $m = 2$ mode, i.e., $0.85(I_2/I_0)_{\rm max}$. 
This position is closely related to the underlying elliptical annulus with the largest intensity difference between the bar and the inter-bar region, which usually varies little with $i$. 
However, the bar-interbar contrast, another method based on the Fourier analysis, has larger uncertainties compared to the Fourier decomposition. 
The most likely explanation is that this method takes FWHM of the bar-interbar contrast radial profile as the bar length, which actually changes significantly with inclination angle.

Another uncertainty in the bar measurement stems from the irregularity of the ellipticity and position angle radial profiles measured from the deprojected images.   
In some mock images, the measured ellipticity profile is flat in the barred region without a clear peak (e.g., bottom panel in Figure~2), making it hard to identify $a_{\rm max}$ or $a_{\rm min}$. 
Thus we take $a_{\rm max}$ as the average value of two radii where the ellipticity decreases to 90\% of the typical value in the flat region. 
$a_{\rm 10}$ may also have problems. 
Model B shows that the measured $e_{\rm 10}$ has a drastic change at different $i$ and $\phi_{\rm bar}$, causing large uncertainties in the 2D analytical deprojection of $a_{\rm 10}$ and $e_{\rm 10}$. 
We try to use $a_{\rm 5}$ ($5^\degree$ position angle deviation) instead of $a_{\rm 10}$, but it does not make a huge difference.

\section{CONCLUSION}

In this work, we use two simulated galaxies to investigate uncertainties of bar deprojection. 
The simulated barred galaxies are projected onto a 2D plane with different bar orientations and disk inclinations. 
The bar properties are measured with three different tracers, i.e., the maximum ellipticity ($a_{\rm max}$, $e_{\rm max}$), minimum ellipticity ($a_{\rm min}$, $e_{\rm min}$), and $10^\degree$ position angle variation ($a_{\rm 10}$, $e_{\rm 10}$). 
Comparing the deprojected parameters with the intrinsic face-on values, we find that the uncertainties increase with increasing $i$. 
When $i$ is larger than $60^\degree$, all deprojection methods fail badly. 

Among all the deprojection methods tested here, the 1D analytical deprojection has the largest uncertainties (up to $\sim 100\%$). 
This method assumes that the bar can be treated as a simple straight line segment, which obviously over-simplifies the structure of the bar. 
It is not surprising that it has relatively large errors because the projected major axis of the bar does not coincide with the real one in the face-on view. 
At relatively smaller $i$ ($\leq 60^\degree$), 2D deprojection methods (analytical and image stretching) and Fourier-based methods (Fourier decomposition and bar-interbar contrast) perform reasonably well with uncertainties $\sim10\%$ in both the bar length and ellipticity. 
Different bar measurement methods also show systematic differences in the deprojection uncertainty. 
For $a_{\rm max}$, both the 1D and 2D methods tend to overestimate the intrinsic bar length, whereas no clear trend can be found for $a_{\rm min}$ and $a_{10}$. 
For the ellipticity, as the bar orientation increases from $0^\degree$ to $90^\degree$, the deprojected $e_{\rm max}$ and $e_{\rm min}$ from 2D methods transition from underestimation to overestimation, while the deprojected $e_{10}$ is generally underestimated.
Bar ellipticity starts to have greater errors at lower inclinations as compared to bar length.

Theoretically, deprojection uncertainties stem from two factors. 
The uncertainties caused by the measurement of inclination angle and $\phi_{bar}$ are much smaller compared to the 3D structure of the bar itself. 
We construct two triaxial toy bar models that can reproduce the results of the 1D and 2D analytical deprojections fairly well; it confirms the vertical thickness of the bar as the main source of uncertainties. 
By comparing the projected ellipse of a 3D triaxial bar with that of a planar ellipse, we find that the projected ellipticity difference is $\sim 0.1$ at $\sim 60^\degree$ inclination angle, which increases at larger $i$. 
Indeed, this difference is the fundamental reason for the deprojection uncertainty.

This is the first work performed on simulated disk galaxies to systematically investigate uncertainties of the deprojection methods, which can provide guidelines for the sample selection and error estimation of future statistical researches on barred galaxies. 
However, our models can be further improved.
For example, including a classical bulge may create an even more realistic ellipticity profile.
We will extend our work to more realistic models in the future.

\acknowledgements The research presented here is partially supported by the 973 Program of China under grants No. 2014CB845701, by the National Natural Science Foundation of China under grants No. 11073037, 11333003, 11322326, by the Strategic Priority Research Program ``The Emergence of Cosmological Structures'' (No. XDB09000000) of Chinese Academy of Sciences, and by Shanghai Yangfan Talent Youth Program (No. 14YF1407700). 
ZYL gratefully acknowledges the support of K. C. Wang Education Foundation (Hong Kong). 
We thank Jerry Sellwood for providing us Model B, which makes this work complete.
Hospitality at APCTP during the 7th Korean Astrophysics Workshop is kindly acknowledged.


\bibliography{reference}

\end{document}